\documentclass[preprint,12pt]{elsarticle}
\usepackage{graphicx}
\usepackage{amssymb}
\usepackage{amsmath}

\journal{Nuclear Physics A}

\begin{document}

\begin{frontmatter}

\title{Testing in-medium $\pi N$ dynamics on pionic atoms} 

\author[a]{E.~Friedman\corref{cor1}}
\author[a]{A.~Gal}
\cortext[cor1]{Corresponding author: E. Friedman, elifried@cc.huji.ac.il}
\address[a]{Racah Institute of Physics, The Hebrew University, 91904
Jerusalem, Israel}

\begin{abstract}
A general algorithm for handling the energy dependence of meson-nucleon 
amplitudes in the nuclear medium has been recently applied to antikaons and 
to eta mesons. Here we test this approach on $\pi N$ amplitudes in pionic 
atoms where direct comparison can be made with ample experimental results. 
Applying this algorithm to a large-scale fit of 100 pionic-atom data points 
across the periodic table, which also include the `deeply-bound' states 
in Sn and Pb, reaffirms earlier conclusions on the density-dependent 
renormalization of the $\pi N$ threshold isovector amplitude $b_1$, 
or equivalently the renormalization of the pion decay constant $f_{\pi}$ 
in the nuclear medium. 
\end{abstract}

\begin{keyword}
pion-nucleon in-medium interaction, energy dependence, pionic atoms
\end{keyword}

\end{frontmatter}

\section{Dedication} 
\label{sec:ded} 

This contribution is dedicated to the memory of Gerry Brown who has charted 
and shaped up the frontiers of Nuclear Physics for half a century. Gerry 
commissioned our two past reviews on exotic atoms \cite{BFG97,FGa07}, 
realizing the potential of exotic atoms to provide hints and constraints 
on chiral signatures in dense matter. A brief update of our work on exotic 
atoms appeared in the 85th birthday Festschrift in honor of Gerald E Brown, 
{\it From Nuclei to Stars}, published in 2011 \cite{FGa11}. Here we revisit 
pionic atoms, a subject to which he made significant contributions in the 
1970s.

\section{Introduction}
\label{sec:intro}

It was recognised in the early 1970s that the $\bar K$-nucleus interaction
near threshold is determined by the $\bar K$-nucleon scattering amplitude at
subthreshold energies \cite{Wyc71,BTo72,Roo75}. The proximity of the 
$\Lambda(1405)$ resonance makes the $\bar K$-nucleus interaction quite 
sensitive to the proper description of the subthreshold region. An algorithm 
devised to account for the subthreshold energy dependence of the 
meson-nucleon amplitude in evaluating the meson-nucleus strong-interaction 
potential in kaonic atoms and for strongly-bound $\bar K$ mesons was 
formulated recently \cite{CFG11,CFG11a,FGa12,GMa12,FGa13}. It has been 
employed in extensive calculations using several models for $\bar K$-nucleon 
scattering amplitudes that respect the recent SIDDHARTA results for kaonic 
atoms of hydrogen \cite{SID12}. Another application of the same algorithm 
was made very recently in calculations of strongly-bound states of $\eta$ 
mesons in nuclei \cite{FGM13,CFG14}. Here the proximity of the $N^*$(1535) 
resonance enhances the sensitivity to the models used for the $\eta$-nucleon 
interaction and to the way the subthreshold region is handled. For an 
up-to-date overview of applications to in-medium $\bar K$ and $\eta$ mesons, 
see Ref.~\cite{GFB14}. 

In the two examples mentioned above no direct comparison with experiment 
could be made. Although there exists a reasonably broad database of strong 
interaction effects in kaonic atoms, the above approach is limited to $K^-$ 
interactions with a single in-medium nucleon. Comparison with experiments 
gave evidence for a sizable component of the $K^-$-nucleus potential which 
represents multinucleon interactions \cite{FGa12,FGa13} thus limiting one's 
ability to directly confront the model with experiment. Similarly, there is 
no experimental information on strongly bound antikaons in nuclei, nor any 
accepted direct evidence for bound $\eta$ nuclear states. 

Pion-nucleus interactions near threshold present a different scenario with 
respect to the above examples. No resonances are known near threshold and the 
real part of the pion-nucleus potential, which is part of the subthreshold 
algorithm, is dominated by single-nucleon interaction terms closely 
approximated by their free-space counterparts. Relatively small contributions 
arise from absorption on two nucleons which gives rise to the imaginary part 
of the potential. Moreover, the experimental database for pionic atoms is 
the most extensive of all exotic atoms \cite{BFG97,FGa07}. It is natural 
then to test the subthreshold approach by analysing pionic atom data where 
direct comparison with experiment is meaningful. On the other hand, owing to 
the rather smooth energy dependence of the free-space $\pi N$ interaction, 
such tests may not prove sufficiently sensitive to the fine details of the 
in-medium $\pi N$ interaction even though they do provide acceptable support 
for the validity of the algorithm. It is therefore encouraging that the fits 
to pionic atom data reported here do preserve and even somewhat improve the 
existing level of agreement with the data, reaffirming earlier conclusions 
on the in-medium renormalization of the $\pi N$ interaction \cite
{KYa01,Fri02,GGG02,YHi03,KKW03,KKW03c,FGa03c,FGa03,FGa04,Suz04,Fri04,Fri05}. 

In Section \ref{sec:model} we present the model behind the algorithm for 
handling subthreshold pion-nucleon amplitudes in the medium. Results of global 
fits to pionic atom data are presented in Section \ref{sec:res}, 
based on empirical energy-dependence of the pion-nucleon amplitudes. 
The in-medium enhancement of the $\pi N$ isovector $s$-wave amplitude 
$b_1$ that received attention in the last two decades is examined and the 
role of the $E \to E-V_c$ `minimal substitution' requirement is discussed. 
A summary and discussion are given in Section \ref{sec:summ}.

\section{Subthreshold model} 
\label{sec:model} 

Here we briefly review the methodology of hadronic-atom calculations, using 
energy-dependent optical potentials within a suitably constructed subthreshold 
model. 

\subsection{Wave equation} 
\label{sec:KG} 

Starting from a meson Klein-Gordon (KG) equation, the dispersion relation 
corresponding to a meson of mass $m$, momentum $\vec p$ and energy $E$ in 
nuclear medium of density $\rho$ is traditionally written as \cite{FGa07} 
\begin{equation}\label{eq:disp} 
E^2-{\vec p}^{~2}-m^2-\Pi(E,\vec p,\rho)=0,   
\end{equation} 
in terms of a self energy (polarization operator) $\Pi(E,\vec p,\rho)$. 
Introducing the meson-nuclear optical potential $V_{\rm opt}$ by 
$\Pi(E,\vec p,\rho)=2EV_{\rm opt}$, the following KG equation is obtained 
for finite nuclei and at or near threshold: 
\begin{equation}\label{eq:KG} 
\left[ \nabla^2  - 2{\mu}(B+V_{\rm opt} + V_c) + (V_c+B)^2\right] \psi = 0, 
\end{equation} 
where $\hbar = c = 1$ was implicitly assumed in these equations. 
Here $\mu$ is the meson-nucleus reduced mass, $B$ is the complex binding 
energy, $V_c$ is the finite-size Coulomb interaction of the meson with 
the nucleus, including vacuum-polarization terms, all added according 
to the minimal substitution principle $E \to E - V_c$. $V_{\rm opt}$ 
is the optical potential of the meson in the medium. Additional terms 
$2V_cV_{\rm opt}$ and $2BV_{\rm opt}$ were neglected in Eq.~(\ref{eq:KG}) 
with respect to $2{\mu}V_{\rm opt}$. 

For pions, the emphasis in past studies of medium effects has been on the 
density dependence of the self energy operator $\Pi(E,\vec p,\rho)$, or 
equivalently of the optical potential $V_{\rm opt}$. The density dependence 
of $V_{\rm opt}$ is briefly reviewed in Section~\ref{sec:potl} below, while 
the energy and momentum dependence, which is the focus of the present work, 
is studied in Section~\ref{sec:Edep}. 

\subsection{Pion-nucleus potential} 
\label{sec:potl} 

The zeroth-order optical potential $V_{\rm opt}^{(0)}=t\rho$ is linear in 
the density $\rho$, with $t$ the free-space meson-nucleon $t$ matrix. This 
optical potential is real at zero energy because pions cannot be absorbed 
at rest by a single nucleon, although they can be absorbed by the nucleus. 
Ericson and Ericson \cite{EEr66} introduced $\rho^2$ terms into the 
potential to describe schematically $\pi^-$ absorption on pairs of 
nucleons. The $\pi^-$ optical potential then becomes 
\begin{equation} \label{eq:EE1} 
2\mu V_{\rm opt}(r) = q(r) + \vec \nabla \cdot \alpha(r) \vec \nabla, 
\end{equation} 
\noindent 
with its $s$-wave part $q(r)$ given by 
\begin{eqnarray} \label{eq:EE1s}
q(r) & = & -4\pi(1+\frac{\mu}{m_N})\{\overline{b}_0[\rho_n(r)+\rho_p(r)]
  +\overline{b}_1[\rho_n(r)-\rho_p(r)] \} \nonumber \\
 & &  -4\pi(1+\frac{\mu}{2m_N})4B_0\rho_n(r) \rho_p(r),
\end{eqnarray}
\noindent
where $\rho_n$ and $\rho_p$ are the neutron and proton density 
distributions normalized to the number of neutrons $N$ and number 
of protons $Z$, respectively, and $m_N$ is the mass of the nucleon. 
The $p$-wave part of the potential, $\alpha(r)$, is effective only 
near the nuclear surface and will not be further discussed here; 
it is used in the present work in the same way as in Ref.~\cite{FGa07}. 
The coefficients $\overline{b}_0$ and $\overline{b}_1$ are {\it effective}, 
implicitly density-dependent pion-nucleon isoscalar and isovector 
$s$-wave scattering amplitudes evolving from the free-space amplitudes 
$b_0$ and $b_1$, respectively, and are essentially real near threshold. 
The parameter $B_0$ represents multi-nucleon absorption and therefore 
has an imaginary part. The real part of $B_0$ stands for dispersive 
contributions which could play a role in pionic atoms. 

The free-space values of the isoscalar and isovector $\pi N$ center-of-mass 
(cm) scattering amplitudes at threshold are known from pionic hydrogen and 
deuterium X-ray measurements \cite{Sch01} 
\begin{equation} \label{eq:Schroder} 
b_0=-0.0001^{+0.0009}_{-0.0021}\;m_{\pi}^{-1}, \;\;\;\;\; 
b_1=-0.0885^{+0.0010}_{-0.0021}\;m_{\pi}^{-1}.  
\end{equation} 
A recent chiral perturbation calculation of the $\pi^-d$ scattering 
amplitude, reanalysing these X-ray measurements, finds the following 
values~\cite{BHH11}: 
\begin{equation} \label{eq:Baru} 
b_0=0.0076\pm 0.0031\;m_{\pi}^{-1}, \;\;\;\;\; 
b_1=-0.0861\pm 0.0009\;m_{\pi}^{-1}. 
\end{equation} 
One notes the close agreement between the two determinations of $b_1$ whereas 
the values of $b_0$, accepting the quoted uncertainties, disagree with each 
other. Yet, these values of $b_0$ and $b_1$ are well approximated by the 
Tomozawa-Weinberg (TW) lowest-order chiral limit \cite{TWe66}
\begin{equation} \label{eq:TW} 
b_0=0,\;\;\;\;\; b_1=-\frac{\mu_{\pi N}}{8 \pi f^{2}_{\pi}}=
-0.079\;m_{\pi}^{-1}. 
\end{equation} 
Since the value of the amplitude $b_0$ from 
Eqs.~(\ref{eq:Schroder})-(\ref{eq:Baru}) is exceptionally small, it is 
mandatory to include double-scattering contributions in the construction 
of the isoscalar single-nucleon term in $q(r)$, giving rise to explicit 
density dependence for Pauli correlated nucleons \cite{KEr69} 
\begin{equation} \label{eq:b0b}
\overline{b}_0 = b_0 - \frac{3}{2\pi}(b_0^2+2b_1^2)p_F,
\end{equation} 
where $p_F$ is the local Fermi momentum corresponding to the local nuclear 
density $\rho=2p_F^3/(3\pi^2)$. Here $b_1^2$ contributes significantly to 
$\overline{b}_0$, thereby making the isovector amplitude $b_1$ effective 
also in $N=Z$ pionic atoms. The TW expression (\ref{eq:TW}) for $b_1$ 
suggests that its in-medium renormalization is directly connected to that 
of the pion decay constant $f_{\pi}$, given to first order in the nuclear 
density $\rho$ by \cite{Wei00,Wei01} 
\begin{equation} \label{eq:fpi} 
\frac{f_\pi^2(\rho)}{f_\pi^2} = \frac{<\bar q q>_{\rho}}{<\bar q q>_0} 
\simeq {1 - \frac{\sigma \rho}{m_{\pi}^2 f_{\pi}^2}}, 
\end{equation} 
where $<\bar q q>_{\rho}$ stands for the in-medium chiral condensate and 
$\sigma\simeq 50$~MeV is the pion-nucleon $\sigma $ term.
Extending Eq.~(\ref{eq:TW}) for $b_1$ from free space to dense matter, 
this leads to the following density dependence for the in-medium $b_1$: 
\begin{equation} \label{eq:ddb1} 
b_{1}(\rho)=\frac{b_{1}}{1 - {\sigma \rho} / {m_{\pi}^2 f_{\pi}^2}}
=\frac{b_1}{1-2.3\rho}  
\end{equation} 
with $\rho$ in units of fm$^{-3}$. In this model, introduced first by 
Weise \cite{Wei00,Wei01}, the explicitly density-dependent $b_{1}(\rho)$ 
of Eq.~(\ref{eq:ddb1}) replaces the effective isovector coefficient 
$\overline{b}_1$ in the pion-nucleus $s$-wave potential $q(r)$ of 
Eq.~(\ref{eq:EE1s}). This has been referred to in the literature as 
a necessary signature of partial restoration of chiral symmetry in 
dense matter. We note that the linear density approximation introduced 
in Eq.~(\ref{eq:fpi}) provides an excellent representation of quark 
condensate effects at densities below nuclear matter density \cite{KHW08}. 

\subsection{Nuclear densities} 
\label{sec:rho}

An important ingredient in the analysis of pionic atoms are the nuclear 
densities to be used in the potential~(\ref{eq:EE1}). With proton 
densities considered known we scan over neutron densities searching for 
a best agreement with the data. A linear dependence of $r_n-r_p$, the 
difference between the root-mean-square (rms) radii, on the neutron excess 
ratio $(N-Z)/A$ has been accepted as a useful quantity, parameterized as
\begin{equation} \label{eq:rnrp} 
r_n-r_p = \gamma \frac{N-Z}{A} + \delta \; ,
\end{equation} 
with $\gamma$ close to 1.0~fm and $\delta$ close to zero. Two-parameter 
Fermi distributions are used for $\rho _n $ with the same diffuseness 
parameter as for the protons, the so-called `skin` shape \cite{FGa07,Fri09} 
which was found to yield lower values of $\chi ^2$ for pions. 

\subsection{Energy dependence}
\label{sec:Edep}

The model underlying the subthreshold algorithm adopts the Mandelstam 
variable $s = (E_{\pi} + E_N)^2 -(\vec p _{\pi} + \vec p_N)^2$ as the 
argument transforming free-space to in-medium pion-nucleon amplitudes, 
where both the pion and the nucleon variables are determined independently 
by the respective environment of a pionic atom and a nucleus. Consequently, 
unlike in the two-body cm system, here $\vec p _{\pi} + \vec p_N$ does not 
vanish, and one gets $(\vec p _{\pi}+\vec p_N)^2\rightarrow p _{\pi}^2+p_N^2$ 
upon averaging over angles. The energies are given by
\begin{equation}
\label{eq:energies}
E_{\pi} =m_{\pi} -B_{\pi},~~~~ E_N=m_N-B_N, 
\end{equation}
where $B$ are binding energies and $m$ are masses. For the pion momentum we 
substitute locally
\begin{equation}
\label{eq:locpi}
\frac{p_{\pi}^2}{2 m_{\pi}} = -B_{\pi} - {\rm Re}~V_{\rm opt} -V_c
\end{equation}
with $V_{\rm opt}$ the pion-nucleus optical potential and $V_c$ the pion 
Coulomb potential due to the finite-size nuclear charge distribution. 
For the nucleon we adopt the Fermi gas model (FGM), yielding in the local 
density approximation 
\begin{equation}
\label{eq:Fermi}
\frac{p_N^2}{2 m_N} = T_N\; (\rho / {\bar\rho})^{2/3}, 
\end{equation}
where $\rho$ is the local density, $\bar\rho$ the average nuclear density and 
$T_N$ is the average nucleon kinetic energy which assumes the value 23~MeV in 
the FGM. This value, used here, is larger than that derived from shell-model 
calculations across the periodic table (18.4 MeV in $^{208}$Pb), but smaller 
than that derived in many-body approaches (38.2 MeV in $^{208}$Pb), 
see Tables I and II in Ref.~\cite{PRC96}. Most of the difference between 
these two groups of values is believed to arise from short-range nuclear 
correlations at the high density regime which is inaccessible in exotic atoms. 
For this reason, furthermore, we kept only the leading density power 2/3 in 
Eq.~(\ref{eq:Fermi}). 

Defining $\delta \sqrt s =\sqrt s -E_{\rm th}$ with $E_{\rm th}=m_{\pi}+m_N$, 
then to first order in $B/E_{\rm th}$ and $(p/E_{\rm th})^2$ one gets 
\begin{equation} 
\delta \sqrt s = -B_N\rho/{\bar \rho} 
-\xi_N [T_N(\rho/\bar{\rho})^{2/3}+B_{\pi}\rho/\rho_0] 
 +\xi_{\pi}[{\rm Re}~V_{\rm opt}+V_c (\rho/\rho_0)^{1/3}], 
\label{eq:SCmodif} 
\end{equation}
with $\xi_N=m_N/(m_N+m_{\pi}),~\xi_{\pi} = m_{\pi}/(m_N+m_{\pi})$, 
and $\rho_0=0.17$~fm$^{-3}$. Following previous applications 
\cite{CFG11,CFG11a,FGa12,GMa12,FGa13} an average binding energy value of 
$B_N=8.5$~MeV is used. The specific $\rho/\rho_0$ and $\rho/{\bar \rho}$ 
forms of density dependence ensure that $\delta\sqrt{s}\rightarrow 0$ when 
$\rho \rightarrow 0$~\cite{FGa13}. 

Another variant of Eq.~(\ref{eq:SCmodif}) is obtained when considering the 
minimal substitution requirement, the importance of which for incorporating 
electromagnetism in a gauge-invariant way into the pion optical potential 
was first pointed out by Ericson and Tauscher~\cite{ET82} and more recently 
emphasized by Kolomeitsev, Kaiser and Weise~\cite{KKW03}. Indeed, the 
application of minimal substitution has been successful in analyses of pionic 
atoms \cite{FGa07,FGa04} and pion scattering at low energies \cite{Fri05}. 
Here $E=E_{\pi}+E_N$ is replaced by $E-V_c$ and then Eq.~(\ref{eq:SCmodif}) 
becomes
\begin{equation} 
\delta \sqrt s = 
-B_N\rho/{\bar \rho} 
-\xi_N [T_N(\rho/\bar{\rho})^{2/3}+B_{\pi}\rho/\rho_0+V_c (\rho/\rho_0)^{1/3}] 
 +\xi_{\pi}{\rm Re}~V_{\rm opt}.  
\label{eq:SCmodifMS} 
\end{equation}
Equations (\ref{eq:SCmodif})-(\ref{eq:SCmodifMS}) show that the 
subthreshold energies where the $\pi N$ amplitudes in $V_{\rm opt}$ are to 
be evaluated depend, locally, on the optical potential $V_{\rm opt}$ which, 
according to Eqs.~(\ref{eq:EE1})-(\ref{eq:b0b}), depends on these amplitudes. 
Therefore these equations have to be solved {\it self-consistently}, 
a process that is found to converge after 4--6 iterations which are done 
at every radial point for the local value of the nuclear density $\rho$. 
This process defines a density-to-energy transformation which makes the 
potential density dependent, in addition to any genuine density dependence 
of the $\pi N$ amplitudes. Only the $s$-wave term of the potential, $q(r)$, 
is used by us to calculate the subthreshold energies in pionic atoms, as 
the $p$-wave term, $\alpha(r)$, is effective primarily near the nuclear 
surface.   

\begin{figure}[htb]
\begin{center}
\includegraphics[height=6cm,width=0.49\linewidth]{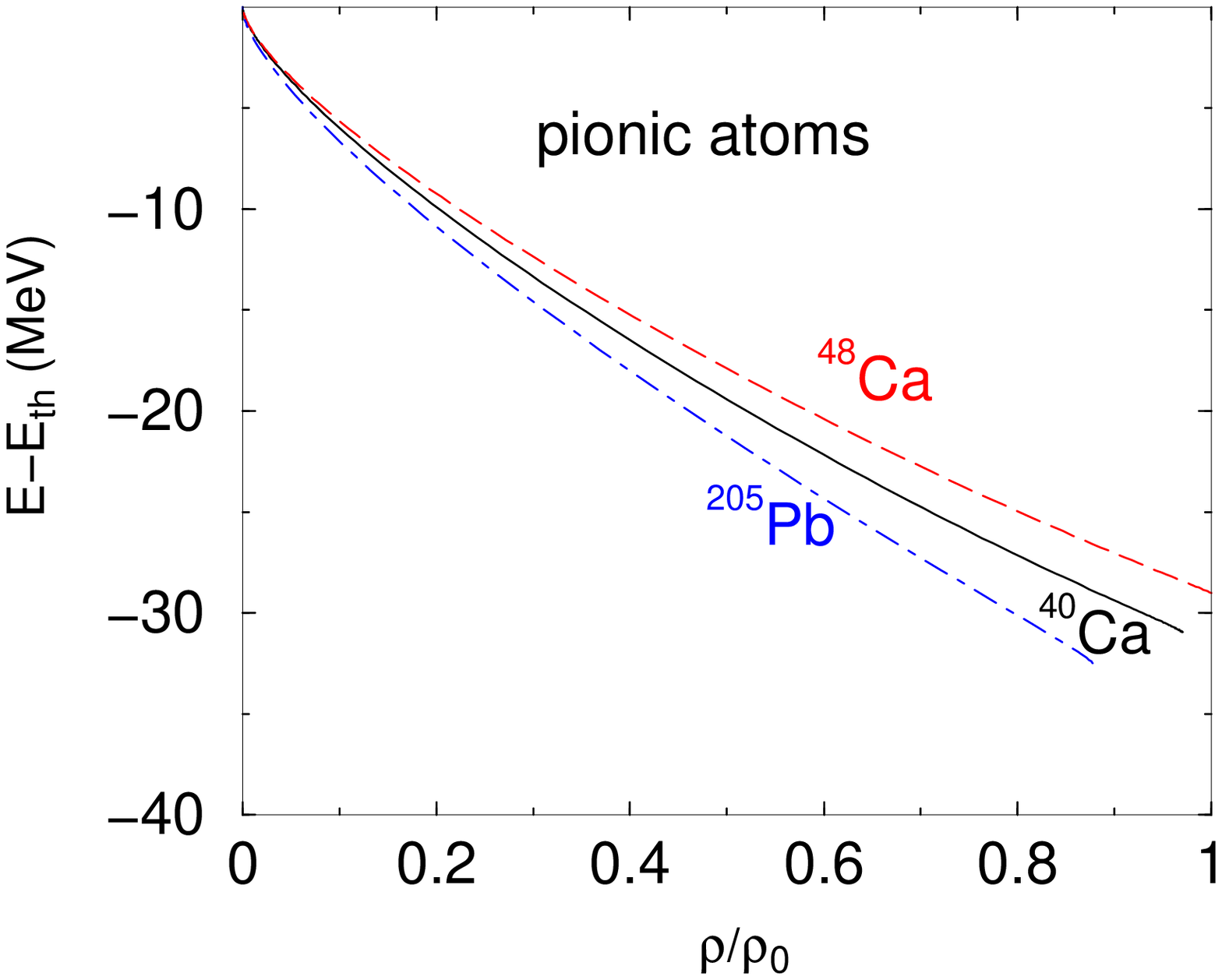}
\includegraphics[height=5.8cm,width=0.49\linewidth]{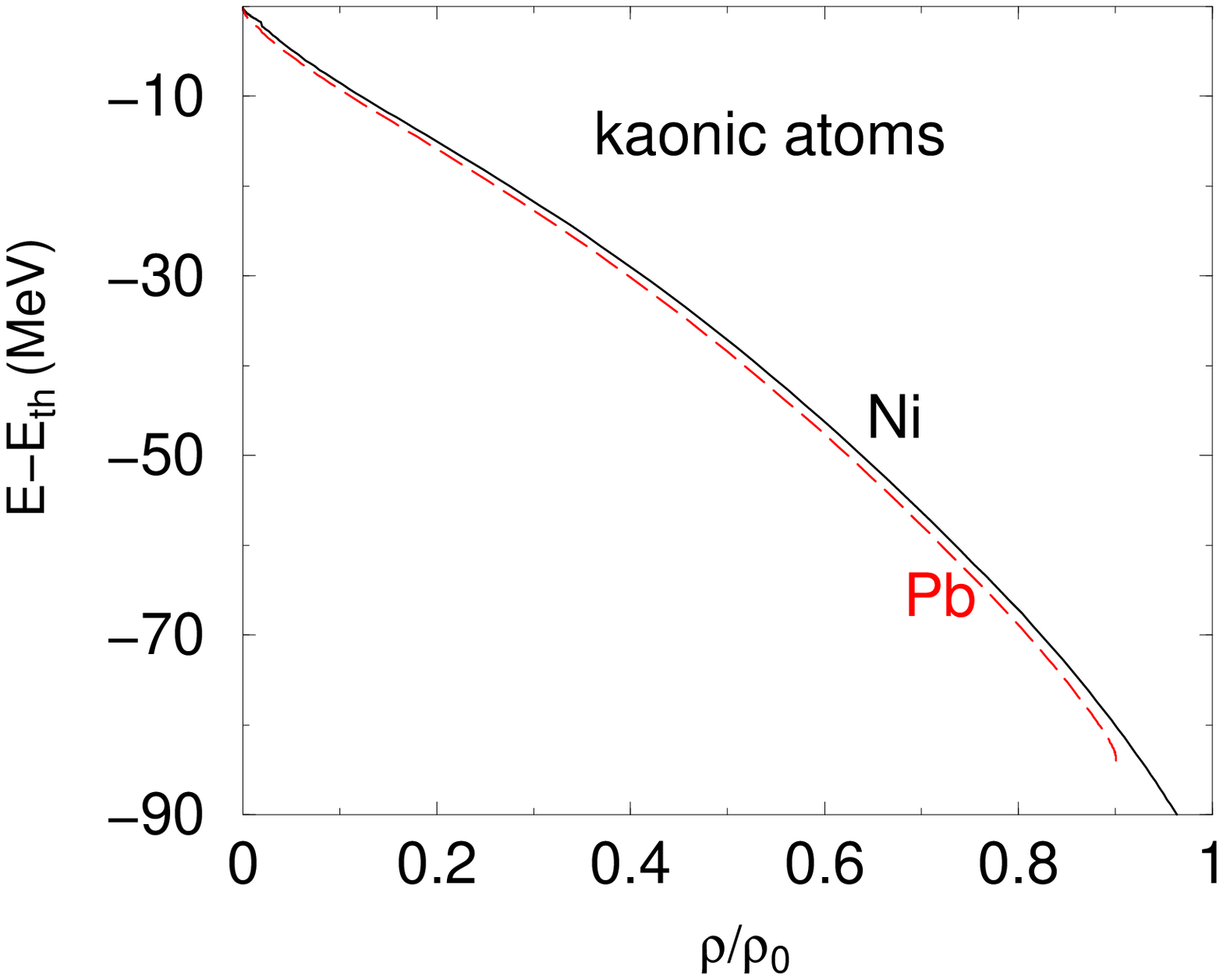} 
\caption{Examples of the density-to-energy transformation in pionic atoms 
(left panel) and in kaonic atoms (right panel, from \cite{FGa13}) using 
Eq.~(\ref{eq:SCmodif}) for pionic atoms and its obvious adjustment for kaonic 
atoms.} 
\label{fig:deltaE} 
\end{center} 
\end{figure} 

Figure~\ref{fig:deltaE} shows examples of the density-to-energy transformation 
for three pionic atoms in comparison with typical results for kaonic atoms, 
with $\delta \sqrt s$ denoted by $E-E_{\rm th}$. Note that the example 
of pionic atom of $^{205}$Pb is an extreme case as it relates to a deeply 
bound $2p$ pionic state with a very large binding energy of close to 6 MeV. 
The empirical $\pi N$ scattering amplitudes used, including their energy 
dependence, were taken from $\pi N$ phase shifts derived from the SAID 
program \cite{SAID}, yielding a slope of $-$0.00053$m_{\pi}^{-1}$/MeV 
for $b_0$. We use this slope also below threshold. For $b_1$, 
its variation with energy is negligibly small and therefore ignored here. 
The figure demonstrates clearly that subthreshold energies of order $-$20 MeV, 
at $\rho_0/2$, are encountered in pionic atoms (left panel). Although smaller 
than typically $-$40 MeV in kaonic atoms (right panel), this constitutes 
a nonnegligible effect. The difference between pionic and kaonic atoms 
originates from the very different meson-nuclear $s$-wave potentials for 
these two systems. For $K^-$, with attractive real part of the potential, 
the energy shift $\delta \sqrt s$ is negative definite when disregarding 
the minimal substitution contribution. For pions, the real part of the 
$s$-wave potential is repulsive but, nevertheless, $\delta \sqrt s$ is 
negative due to the dominance of the $-p^2$ contributions in the expression 
for the Mandelstam energy variable $s$. The significance of these 
momentum-dependent contributions precludes relating to pions in finite 
nuclei as zero-momentum pions, a term used often for pions in nuclear matter.

\section{Results}
\label{sec:res}

In studying the model for handling pion-nucleon interactions in the nuclear 
medium we compare our new results with earlier analyses \cite{FGa07} where 
the interactions were either taken at threshold or effectively above 
threshold when adopting the minimal substitution principle. We therefore 
consider here, as in Ref.~\cite{FGa07}, global fits to 100 data points for 
strong interaction observables in pionic atoms, from Ne to U, including 
deeply bound pionic atoms of three Sn isotopes and of $^{205}$Pb. 

\begin{figure}[!h] 
\begin{center} 
\includegraphics[height=8cm,width=0.7\linewidth]{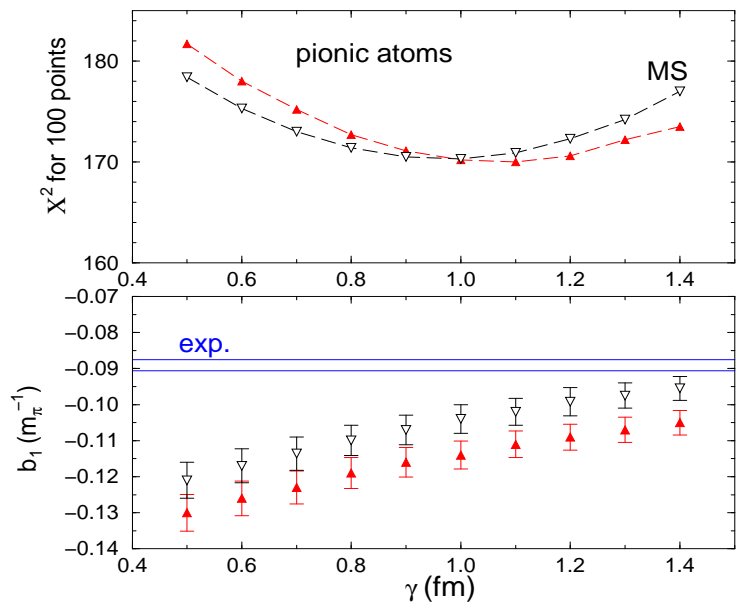} 
\caption{Results of fits for variable $\gamma $, the neutron rms radius 
parameter in Eq.~(\ref{eq:rnrp}), with $b_1$ not modified in the medium. 
Filled triangles up: $\delta \sqrt s$ from Eq.~(\ref{eq:SCmodif}), opaque 
triangles down: including also minimal substitution (MS) as per 
Eq.~(\ref{eq:SCmodifMS}). Also shown is the experimental value of 
$b_1$ at threshold \cite{Sch01}.} 
\label{fig:b1noW} 
\end{center} 
\end{figure} 

The ability to perform least-squares fits to a broad data set for 
pionic atoms is a major difference between the present work and our 
earlier applications of the sub-threshold approach. For example, changing 
input parameters such as $T_N$ in Eqs.~(\ref{eq:SCmodif})-(\ref{eq:SCmodifMS}) 
causes values of fit parameters to change too in order to maintain best 
agreement with the data. Replacing $\bar \rho$ by $\rho _0$ in the $T_N$ 
term, for example, results in the parameter $b_0$ changing within error, 
to compensate for changes due to the $T_N$ term. Most important is the 
observation that changes in the isovector parameter $b_1$ are only 30\% 
of their (rather small) uncertainties. 

Figure \ref{fig:b1noW} shows typical results of fits obtained when varying 
parameters of the pion-nucleus potential and scanning over the neutron rms 
radius parameter $\gamma$, for a fixed value of $\delta=-0.035$~fm, see 
Eq.~(\ref{eq:rnrp}). Finite-range folding in the $p$-wave term is included 
as before \cite{FGa07}. A shallow minimum determines $\gamma$ to be around 
1.0$\pm$0.15~fm, in full agreement with a multitude of other 
results \cite{Fri09,JTL04}. For a check, we note that the value obtained 
from Eq.~(\ref{eq:rnrp}) for $^{208}$Pb is $r_n-r_p$=0.18$\pm$0.03~fm, 
in agreement with more direct recent derivations \cite{Fri12,MAMI13}. 

\begin{table}[hbt]
\begin{center}
\caption{Values of $B_0$ ($m_{\pi}^{-4}$ units) obtained in large-scale 
pionic-atom fits, with minimal substitution (MS) disregarded (No) or 
employed (Yes), using the subthreshold energy algorithms~(\ref{eq:SCmodif}) 
and (\ref{eq:SCmodifMS}) respectively, with (Yes) or without (No) Weise's 
renormalization ansatz~(\ref{eq:ddb1}).}  
\begin{tabular}{ccccc} 
\hline 
MS/Weise & No/No & No/Yes & Yes/No & Yes/Yes \\ 
\hline 
$-$Re$\;B_0$ & 0.112$\pm$0.031 & 0.040$\pm$0.032 & 0.076$\pm$0.032 & 
0.012$\pm$0.033 \\ 
$~$Im$\;B_0$ & 0.052$\pm$0.002 & 0.051$\pm$0.002 & 0.052$\pm$0.002 & 
0.052$\pm$0.002 \\ 
\hline
\end{tabular}
\label{tab:B_0}
\end{center}
\end{table}

\begin{figure}[!ht] 
\begin{center} 
\includegraphics[height=8cm,width=0.7\linewidth]{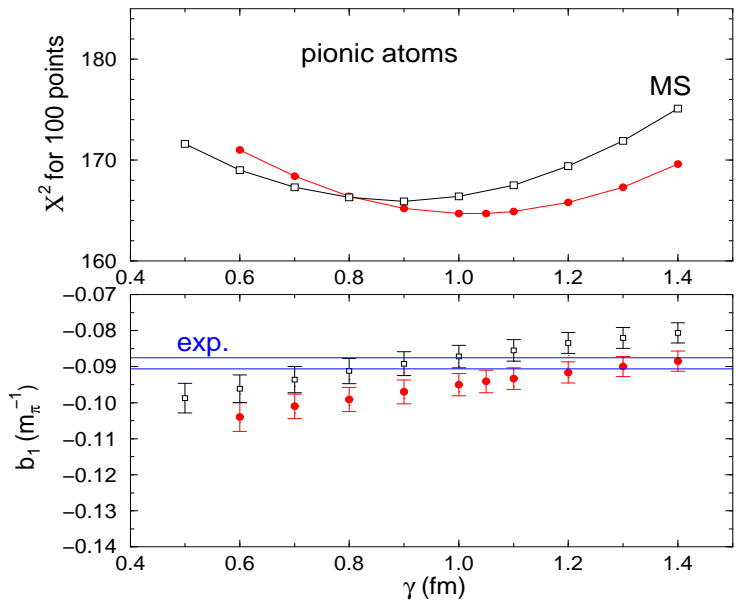} 
\caption{Results of fits for variable $\gamma $, the neutron 
rms radius parameter in Eq.~(\ref{eq:rnrp}), with $b_1(\rho )$ 
given by Eq.~(\ref{eq:ddb1}). Filled: $\delta \sqrt s$ from 
Eq.~(\ref{eq:SCmodif}), opaque: including also minimal substitution 
(MS) as per Eq.~(\ref{eq:SCmodifMS}). The free-space threshold value 
of $b_1$ \cite{Sch01} is also shown.} 
\label{fig:b1W} 
\end{center} 
\end{figure} 

Focusing on the isovector $s$-wave parameter $b_1$, it is seen from 
Fig.~\ref{fig:b1noW} that when it is not modified in the medium, other 
than by the subthreshold energy algorithm~(\ref{eq:SCmodif}), it turns 
out to be too repulsive by 5-6 error bars compared to the free pion-nucleon 
experimental value. When minimal substitution is imposed as per 
Eq.~(\ref{eq:SCmodifMS}), the discrepancy becomes slightly smaller but, 
nevertheless, the so-called $b_1$ anomalous enhancement is observed. 
These conclusions are in agreement with earlier work not using the 
subthreshold energy algorithm. 

The possibility that $b_1$ is enhanced in the nuclear medium was put 
forward by Weise \cite{Wei00,Wei01}, see Eq.~(\ref{eq:ddb1}). This ansatz 
is supported by experiments both below threshold \cite{KYa01,Fri02} and 
above \cite{Fri04,Fri05}. Figure \ref{fig:b1W} shows that the best $\chi ^2$ 
is now 4-5 units smaller than without this explicit density dependence of 
$b_1(\rho)$ and the values of the free-space $b_1$ from the fit agree with 
the known free value, with a hint that the agreement is better when minimal 
substitution is applied. Another indication in favor of combining the 
renormalization ansatz~(\ref{eq:ddb1}) together with minimal substitution 
$E\to E-V_c$ is provided by inspecting the values obtained in our large-scale 
pionic-atom fits for the two-nucleon parameter $B_0$ of Eq.~(\ref{eq:EE1s}), 
as shown in Table~\ref{tab:B_0}. It is seen that the value of Im$\;B_0$ is 
determined in great precision and is insensitive to the variants used. In 
contrast, the fitted value of Re$\;B_0$ is sensitive to these variants and it 
also emerges with a fairly large uncertainty which, however, is independent 
of the variant used. Anticipating values of Re$\;B_0$ smaller or equal in 
magnitude to Im$\;B_0$, the fitted values listed in the table give preference 
to accepting both the minimal-substitution principle and Weise's 
renormalization ansatz \cite{Wei00,Wei01}.

\section{Summary and discussion} 
\label{sec:summ} 

We have presented in this update global analyses of strong interaction 
effects in pionic atoms, where the algorithm applied recently \cite{GFB14} 
in other systems for handling meson-nucleon in-medium interactions near 
threshold is tested. This approach defines energy scale 
of $\approx$20~MeV over which the $\pi N$ $s$-wave amplitudes contribute to 
the pion-nucleus interaction. These energies are below threshold when minimal 
substitution is disregarded, but part of the energy range can be above 
threshold when minimal substitution is maintained. In all previous analyses, 
before introducing these algorithms, the $\pi N$ energy was either on 
threshold when minimal substitution was disregarded, or fully above threshold 
when respecting minimal substitution. Nevertheless, the overall conclusions 
are practically the same whether or not the present algorithms are applied. 
In particular, the lowest $\chi ^2$ implies a value of $\chi ^2$ per degree 
of freedom of about 1.7 and is obtained for rms radii of neutron distributions 
that agree with those deduced by other methods. The best-fit $b_1$ values 
agree with the free pion-nucleon value only when explicit density-dependence 
of $b_1$ as suggested by Weise \cite{Wei00,Wei01} is incorporated. 
These results were obtained using the SAID empirical energy dependence 
of $b_0$, as distinct from the chiral-model energy dependence evaluated 
for zero-momentum pions. We refer the reader to our earlier discussion 
of this point \cite{FGa04} adding that pions have appreciable momentum 
within the nuclear interior, as given by Eq.~(\ref{eq:locpi}). 

The low sensitivity revealed in the present work results from the smooth 
dependence on energy of the empirical $b_0$ parameter. Considering the energy 
at which $b_0$ is calculated, and assuming an average effective density of 
50-60\% of the central density \cite{GGG02,YHi03}, then the respective energy 
differences between any previous analysis and its present counterpart are 
around 20~MeV. With an empirical $b_0$ slope of $-$0.00053$m_{\pi}^{-1}$/MeV 
\cite{SAID} a variation of 0.01~$m_{\pi}^{-1}$ in $b_0$ values is implied, 
which is indeed produced in the respective $\chi ^2$ fits. Furthermore, with 
$b_1$ essentially independent of energy, the variations in $b_0$ translate to 
these same variations in ${\overline b}_0$. This $\approx$0.01~$m_{\pi}^{-1}$ 
variation in $b_0$ is also consistent with the uncertainty placed on it in 
the SAID partial-wave analysis~\cite{SAID}. This somewhat trivial result is, 
nevertheless, reassuring in supporting the viability of the subthreshold 
algorithm.

Finally we comment that the results and conclusions reached in the present 
update hold valid also when the `deeply bound' states (DBS) in $^{205}$Pb 
and in $^{115,119,123}$Sn are removed from the large-scale fits of pionic 
atoms described here. This agrees with earlier conclusions reached by 
us \cite{FGa11,FGa03c,FGa03} where it was shown that the uncertainties in 
the derived value of $b_1$ from DBS, or from DBS plus $1s$ states in light 
$N=Z$ nuclei, are considerably larger than those derived from large-scale fits 
(about 0.004~$m_{\pi}^{-1}$, see Figs.~\ref{fig:b1noW} and \ref{fig:b1W}). 
Other approaches considering only partial sets of pionic-atom data and 
claiming similar uncertainties, as reviewed in Ref.~\cite{YHH12}, ignore 
additional systematic uncertainties arising from fixing the $p$-wave $\pi N$ 
potential $\alpha(r)$ in Eq.~(\ref{eq:EE1}). We stress that our large-scale 
fits involve a comprehensive parameter search, including the $p$-wave 
interaction parameters not discussed here. The few DBS established accurately 
so far are still short of providing {\it on their own} the necessary precision 
to substantiate the deduced renormalization of $b_1$ and the implied partial 
restoration of chiral symmetry in dense matter through the decrease of 
$f_{\pi}$, given to a good approximation by Eq.~(\ref{eq:fpi}).

\section*{Acknowledgments} 
Special thanks are due to Wolfram Weise for many stimulating discussions 
on exotic atoms throughout the last two decades. This research was partly 
supported by the HadronPhysics3 networks SPHERE and LEANNIS of the European 
FP7 initiative.


\begin{thebibliography}{99}

\bibitem{BFG97} C.J.~Batty, E.~Friedman, A.~Gal, Phys. Rep. 287 (1997) 385. 

\bibitem{FGa07} E.~Friedman, A.~Gal, Phys. Rep. 452 (2007) 89. 

\bibitem{FGa11} E.~Friedman, A.~Gal, in: Sabine Lee (Ed.), 
From Nuclei to Stars, World Scientific, 2011, pp.127-140. 

\bibitem{Wyc71} S.~Wycech, Nucl. Phys. B 28 (1971) 541. 

\bibitem{BTo72} W.A.~Bardeen, E.W.~Torigoe, Phys. Lett. B 38 (1972) 135. 

\bibitem{Roo75} J.R.~Rook, Nucl. Phys. A 249 (1975) 466. 

\bibitem{CFG11} A.~Ciepl\'{y}, E.~Friedman, A.~Gal, D.~Gazda, J.~Mare\v{s},
Phys. Lett. B 702 (2011) 402.

\bibitem{CFG11a} A.~Ciepl\'{y}, E.~Friedman, A.~Gal, D.~Gazda, J.~Mare\v{s},
Phys. Rev. C 84 (2011) 045206.

\bibitem{FGa12} E.~Friedman, A.~Gal, Nucl. Phys. A 881 (2012) 150.

\bibitem{GMa12} D.~Gazda, J.~Mare\v{s}, Nucl. Phys. A 881 (2012) 159.

\bibitem{FGa13} E.~Friedman, A.~Gal, Nucl. Phys. A 899 (2013) 60. 

\bibitem{SID12} M.~Bazzi, et al., Phys. Lett. B 704 (2011) 113, 
Nucl. Phys. A 881 (2012) 88.

\bibitem{FGM13} E.~Friedman, A.~Gal, J.~Mares, Phys. Lett. B 725 (2013) 334.

\bibitem{CFG14} A.~Ciepl\'{y}, E.~Friedman, A.~Gal, J.~Mare\v{s},
Nucl. Phys. A 925 (2014) 126.

\bibitem{GFB14} A.~Gal, E.~Friedman, N.~Barnea, A.~Ciepl\'{y}, D.~Gazda, 
J.~Mare\v{s}, Acta Phys. Pol. B 45 (2014) 673. 

\bibitem{KYa01} P.~Kienle, T.~Yamazaki, Phys. Lett. B 514 (2001) 1, 
using preliminary results reported subsequently by H.~Geissel, et al., 
Phys. Rev. Lett. 88 (2002) 122301. 

\bibitem{Fri02} E.~Friedman, Phys. Lett. B 524 (2002) 87. 

\bibitem{GGG02} H.~Geissel, et al., Phys. Lett. B 549 (2002) 64. 

\bibitem{YHi03} T.~Yamazaki, S.~Hirenzaki, Phys. Lett. B 557 (2003) 20. 

\bibitem{KKW03} E.E.~Kolomeitsev, N.~Kaiser, W.~Weise, Phys. Rev. Lett. 
90 (2003) 092501. 

\bibitem{KKW03c} E.E.~Kolomeitsev, N.~Kaiser, W.~Weise, Nucl. Phys. A 721 
(2003) 835c.  

\bibitem{FGa03c} E.~Friedman, A.~Gal, Nucl. Phys. A 721 (2003) 842c. 

\bibitem{FGa03} E.~Friedman, A.~Gal, Nucl. Phys. A 724 (2003) 143. 

\bibitem{FGa04} E.~Friedman, A.~Gal, Phys. Lett. B 578 (2004) 85. 

\bibitem{Suz04} K.~Suzuki, et al., Phys. Rev. Lett. 92 (2004) 072302. 

\bibitem{Fri04} E.~Friedman, et al., Phys. Rev. Lett. 93 (2004) 122302.

\bibitem{Fri05} E.~Friedman, et al., Phys. Rev. C 72 (2005) 034609. 

\bibitem{EEr66} M.~Ericson, T.E.O.~Ericson, Ann. Phys. 36 (1966) 323. 

\bibitem{Sch01} H.-Ch.~Schr\"oder, et al., Eur. Phys. J. C 21 (2001) 473. 

\bibitem{BHH11} V.~Baru, C.~Hanhart, M.~Hofenrichter, B.~Kubis, A.~Nogga, 
D.R.~Phillips, Phys. Lett. B 694 (2011) 473. 

\bibitem{TWe66} Y.~Tomozawa, Nuovo Cimento A 46 (1966) 707; S.~Weinberg, 
Phys. Rev. Lett. 17 (1966) 616. 

\bibitem{KEr69} M.~Krell, T.E.O.~Ericson,  Nucl. Phys. B 11 (1969) 521. 

\bibitem{Wei00}W.~Weise, Acta Phys. Pol. B 31 (2000) 2715.

\bibitem{Wei01}W.~Weise, Nucl. Phys. A 690 (2001) 98c. 

\bibitem{KHW08} N.~Kaiser, P.~de~Homont, W.~Weise, Phys. Rev. C 77 (2008) 
025204. 

\bibitem{Fri09} E.~Friedman, Hyperfine Interactions 193 (2009) 33. 

\bibitem{PRC96} C.~Ciofi degli Atti, S.~Simula, Phys. Rev. C 53 (1996) 1689. 

\bibitem{ET82} T.E.O.~Ericson, L.~Tauscher, Phys. Lett. B 112 (1982) 425. 

\bibitem{SAID} SAID program gwdac.phys.gwu.edu, see R.A.~Arndt, W.J.~Briscoe, 
I.I.~Strakovsky, R.L.~Workman, Phys. Rev. C 74 (2006) 045205. 

\bibitem{JTL04} J. Jastrz\c{e}bski, et al., Int. J. Mod. Phys. E 13 
(2004) 343. 

\bibitem{Fri12} E.~Friedman, Nucl. Phys. A 896 (2012) 46. 

\bibitem{MAMI13} C.M.~Tarbert, et al. 
(Crystal Ball at MAMI and A2 Collaboration), arXiv:1311.0168. 

\bibitem{YHH12} T.~Yamazaki, S.~Hirenzaki, R.S.~Hayano, H.~Toki, 
Phys. Rep. 514 (2012) pp. 1-87. 



\end{thebibliography}
\end{document}